\documentclass[prl,twocolumn]{revtex4-1}
\usepackage{graphicx,amsmath}
\usepackage{epstopdf,color}

\def\tn{s}
\def\bq{\mathbf q}

\def\re#1{(\ref{#1})}
\def\f{\frac}

\begin{document}

\title{Guyer-Krumhansl-type heat conduction at room temperature}

\author{V\'an P.$^{123}$, Berezovski, A.$^{4}$, F\"ul\"op T.$^{13}$, Gr\'of Gy.$^{1}$, Kov\'acs R.$^{123}$, Lovas \'A.$^{1}$ and Verh\'as J.$^{3}$}

\address{
$^1$Department of Energy Engineering, BME, Budapest, Hungary; \\
$^2$Department of Theoretical Physics, MTA Wigner Research Centre for Physics, Institute for Particle and Nuclear Physics, Budapest, Hungary;\\ 
$^3$Montavid Thermodynamic Research Group, Budapest, Hungary; \\
$^4$Department of Cybernetics, School of Science, Tallinn University of Technology, Tallinn, Estonia}
\date{\today}

\begin{abstract}
Results of heat pulse experiments in various artificial and natural materials are reported in the paper. The experiments are performed at room temperature with macroscopic samples. It is shown that temperature evolution does not follow the Fourier's law but well explained by the Guyer-Krumhansl equation. The observations confirm the ability of non-equilibrium thermodynamics to formulate universal constitutive relations for thermomechanical processes.
\end{abstract}
\maketitle

Deviation from Fourier's law at low temperatures was predicted by Tisza and Landau  and then observed first by Peshkov in liquid Helium \cite{Tisz+}. The wave like propagation of heat, the second sound, is later explained by a hyperbolic extension of Fourier's law called Maxwell-Cattaneo-Vernotte (MCV) equation \cite{Cat+}. The next important experimental step, the detection of second sound and ballistic propagation in solids  \cite{AckGuy+} required an extension of the theory and a generalization lead to the Guyer-Krumhansl equation \cite{Cal+}, which is intensively studied also nowadays, first of all in case of nanomaterials \cite{SelEta+,Zhu+}.

The Guyer-Krumhansl (GK) equation is as follows:
\begin{equation}
\tau \dot{\bq}+ \bq + \lambda_F \nabla T  - \beta' \Delta \bq - \beta'' \nabla\cdot \nabla \bq= 0.
\label{GK}\end{equation}
Here $\bq$ is the heat flux, the current density of the internal energy in a rest frame of the continuum, $\lambda_F$ is the Fourier heat conduction coefficient, $\tau$ is a relaxation time, while $\beta'$ and $\beta''$ are Guyer-Krumhansl coefficients in isotropic media. In rarefied gases these are related to the relaxation times of the Callaway collision integral \cite{Cal+}. The overdot denotes the time derivative, $\nabla$ is the space derivative, and the coefficients are considered as constant. The first three terms of \re{GK} form the Maxwell--Cattaneo--Vernotte equation, the second and the third ones are Fourier's law. 

The conceptual foundation of looking non-Fourier phenomena in macroscopic samples at room temperature is originated in the phenomenological theories: phonons mean free paths and relaxation times are not relevant concepts in sand or frozen meat at room temperature heat conduction. Therefore, among the many phenomenological ideas leading to and explaining the origin of MCV equation (see e.g. \cite{JosPre,Cim+}) non-equilibrium thermodynamics plays a distinguished role. With a consistent phenomenology the obtained constitutive relations are independent of the microscopic details, like Fourier's law in local equilibrium \cite{GroMaz+}. Their validity is based only on the second law of thermodynamics and in this sense they are {\em universal}. Therefore, after the development of Extended Thermodynamics \cite{Gya+,MulRug98b} there were several predictions of similar phenomena with various assumptions and conditions \cite{JosPre}, including heterogeneous materials at room temperature. The first promising positive experimental results in granular and biological media were reported about Maxwell-Cattaneo-Vernotte type heat conduction \cite{Kam+}. However, these measurements were not confirmed \cite{HerBec+}.

In the framework of Extended Thermodynamics, Guyer-Krumhansl equation can be obtained by a modification of the entropy flux in all approaches, usually together with other assumptions \cite{Rug+}. A minimal functional deviation of the entropy flux from the local equilibrium -- introducing Ny\'iri multipliers \cite{Nyi91a1} -- leads to the GK equation without any further ado. It is also straightforward  to obtain most of the viable suggested generalisations of Fourier's law and MCV equation in a uniform framework \cite{VanFul12a}, which is also compatible with kinetic theory \cite{KovVan15a}.

Motivated by the universality of non-equilibrium thermodynamics, most recently an experimental-theoretical study has been performed in order to identify suitable qualitative signatures of detecting non-Fourier heat conduction beyond the MCV equation in heterogeneous, macroscopic samples at room temperature \cite{CzeEta+}. The first indication of the effect was measured on artificial samples, with alternating layers of good and bad heat conductors parallel to the heat flux \cite{BotEta16a}. Here we report experimental results of heat pulse measurements on various artificial and natural materials with GK-type heat conduction.

In our simple experimental device, a flash lamp generates the heat pulse at the front end of the sample, and the temperature is measured by a pin-thermocouple (K-type) at the rear end. The thermocouple and the detector part are insulated from the heat pulse and from the electromagnetic noises. The heat pulse is measured directly at the front end by a photovoltaic cell, triggering data acquisition. A typical pulse shape is triangular and is $10$ ms long.
\begin{figure}[ht]
	\centering
	\includegraphics[width=6cm]{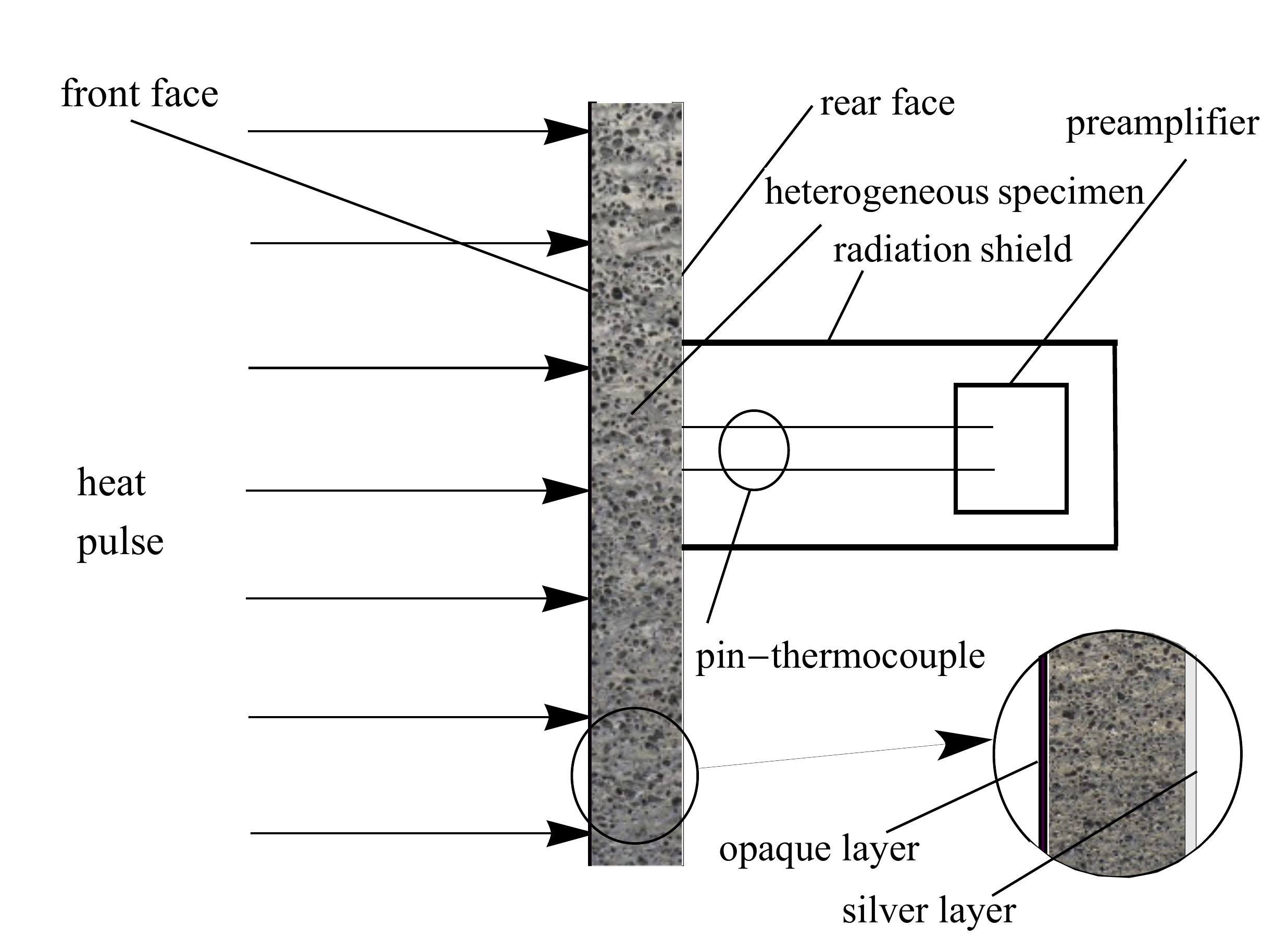}
	\caption{Sketch of the heat pulse experimental setup.}
	\label{fig:exp}
\end{figure}
A sketch of the experimental setup is shown on Figure \ref{fig:exp}. The thickness of the studied specimens is smaller at about a magnitude than their diameter and the front face is homogeneously heated by the heat pulse, therefore, heat conduction can be considered one-dimensional.

At the front side a thin black coating is applied to ensure uniform boundary conditions, as well as to eliminate the transparency of the sample. At the rear side silver painting is used, therefore, the thermocouple measures an effective temperature. The effect of the coating is negligible according to control measurements. The experimental device was calibrated by using several samples with known heat diffusivity.


In a one dimensional setting the balance of internal energy for rigid heat conductors is
\begin{equation}
\rho c \f{\partial T}{\partial t} + \f{\partial q}{\partial x} = 0. \label{ebal}
\end{equation}

Here $c$ is the specific heat, $T$ is the temperature, and the specific internal energy is $e_{sp} = cT$. $\rho$ is the density, and $q$ is the heat flux in the direction of the heat propagation. The one dimensional version of \re{GK} is:
\begin{equation}
\tau \f{\partial q}{\partial t} + q + \lambda_F \f{\partial T}{\partial x} - l^2 \f{\partial^2 q}{\partial x^2} = 0. \label{GK1} 
\end{equation}
Here $k$ is the Fourier heat conduction coefficient, $\tau$ is the relaxation time and $l^2= \beta'+\beta''$  is a nonnegative material parameter of the GK equation, and is expressed with the help of a characteristic length scale $l$. 

Remarkable and instructive to recognise a hierarchical structure in the above system of equations, \re{ebal}--\re{GK1}, when expressed for the temperature \cite{BerEta+}. It is best seen by eliminating the heat flux and rearranging the system as follows:
\begin{equation}
\tau \f{\partial}{\partial t}\left( \f{\partial T}{\partial t} - b\alpha \f{\partial^2 T}{\partial x^2} \right) + \left(\f{\partial T}{\partial t} -\alpha \f{\partial^2 T}{\partial x^2}\right) =0.
\label{hGK} 
\end{equation}
where $\alpha=\f{k}{\rho c}$ is the thermal diffusivity and $b = \f{l^2}{\tau\alpha}$ is the coefficient characterising the deviation from Fourier heat conduction. One can observe that solutions of the Fourier equation are solutions of \re{hGK}, whenever $b=1$, that is $\alpha = \f{l^2}{\tau}$. If  $b< 1$ then the solutions of \re{ebal}--\re{GK1} show wavelike characteristics; if $b>1$ then the solutions are over-diffusive \cite{TanAra00a,KovVan15a}. 

We are looking for solutions of \re{ebal}--\re{GK1} with heat pulse boundary condition and surface heat exchange at the front side. Therefore,
\begin{center}
	$q(x=0,t)= \left\{ \begin{array}{cl}
	q_{\rm max}\left(1-\cos\left(2 \pi \frac{t}{t_p}\right)\right) - B (T(0,t)-T_0) & 
	\textrm{if }\   0<t \leq t_p,\\
	- B (T(0,t)-T_0) & \textrm{if }\ t>t_p.
	\end{array} \right.$
\end{center}
Here $t_p$ is the pulse length, $T_0$ is the ambient temperature and $B$ is the heat exchange coefficient. The particular shape of the pulse does not influence the effect as long as the length of the pulse is much shorter than the characteristic time scale of the experiment. 
The backside boundary is considered adiabatic, $q(L,t)=0$, because of the insulating cover. Initially, the temperature distribution is uniform and the heat flux is zero along the sample, that is, $T(x,0)=T_0$ and $q(x,0)=0$, this is provided by the measurement protocol, with 30-60 minutes temperature equilibration periods between the measurements. 

A suitable dimensionless form of the variables are: 
\begin{gather}
\hat{t} =\frac{t}{t_p}, \quad
\hat{x}=\frac{x}{L}, \quad
\hat{q}=\frac{q}{q_{\rm max}},\quad\hat{T}=\tn\frac{T-T_{0}}{T_{\rm ref}-T_{0}}, \\
q_{\rm max}\! =\! \frac{1}{t_p}  \int_{0}^{t_p}\!\! q_{0}(t)dt, \,
T_{\rm end}=T_{0}+\frac{{q}_{\rm max} t_p}{\rho c L}, 
\tn = \frac{T_{\rm ref}-T_{0}}{T_{\rm end}-T_{0}}, \nonumber
\label{ndvar}\end{gather}
where $q_0(t) = q(x=0,t)+ B(T(0,t)-T_0)$ is the boundary condition without cooling.  $T_{\rm ref}$ is a reference temperature, e.g. the measured maximum temperature, which is different from the limiting adiabatic temperature, $T_{\rm end}$,  because of the cooling. The dimensionless parameters are, consequently,
\begin{equation}
\hat{\tau}    \!=\frac{\tau}{t_p}; \,\,
\hat{\alpha}  \!= \alpha\frac{t_p}{ L^2};  \,\,
\hat{l} 	 \! = \f{l}{L};  \,\,
\hat b \!= \frac{\hat l^2}{\hat \tau\hat\alpha}= b;  \,\,
\hat B \!=B\frac{t_p}{\rho c L}.
\end{equation}

The nondimensional form of the equations is
\begin{eqnarray}
\tn\f{\partial \hat T}{\partial \hat t} + \f{\partial \hat  q}{\partial \hat  x} &=& 0, \label{ndebal}\\
\hat \tau \f{\partial \hat  q}{\partial \hat  t} + \hat  q + \tn\hat  \alpha \f{\partial \hat  T}{\partial \hat x} - b\hat \tau\hat{\alpha} \f{\partial^2 \hat q}{\partial \hat{x}^2} &=& 0. \label{ndGK} 
\end{eqnarray}

The corresponding boundary and initial conditions are:
\begin{center}
	$\hat  q(0,\hat t)= \left\{ \begin{array}{cl}
	1-\cos\left(2 \pi \cdot \hat t\right) - \tn\hat B\hat T(0,t)& 
	\textrm{if } 0<\hat t \leq 1,\\
	- \tn\hat B\hat T(0,t) & \textrm{if } \hat t>1.
	\end{array} \right. $
\end{center}
$\hat  q(1,\hat t)=0$, $\hat T(\hat x,0)=0$ and $\hat q(\hat x,0)=0$. Here the parameters $\tau$, $\alpha$ and $b$ characterise the material, the temperature scale, $\tn$, and the heat exchange coefficient, $B$, are not. 

\begin{figure}[ht]
	\centering
	\includegraphics[width=2cm,height=2cm]{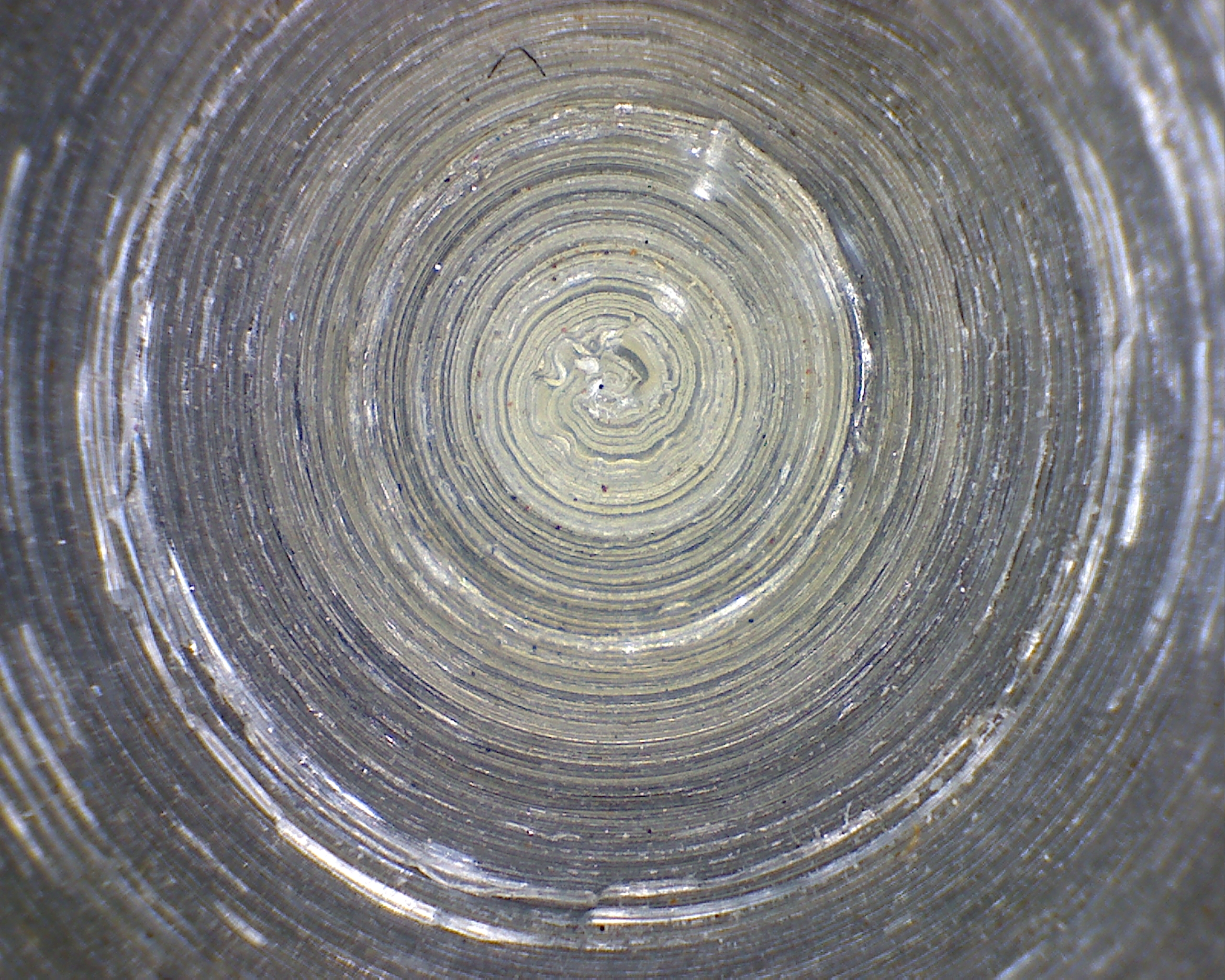}
	\includegraphics[width=2cm,height=2cm]{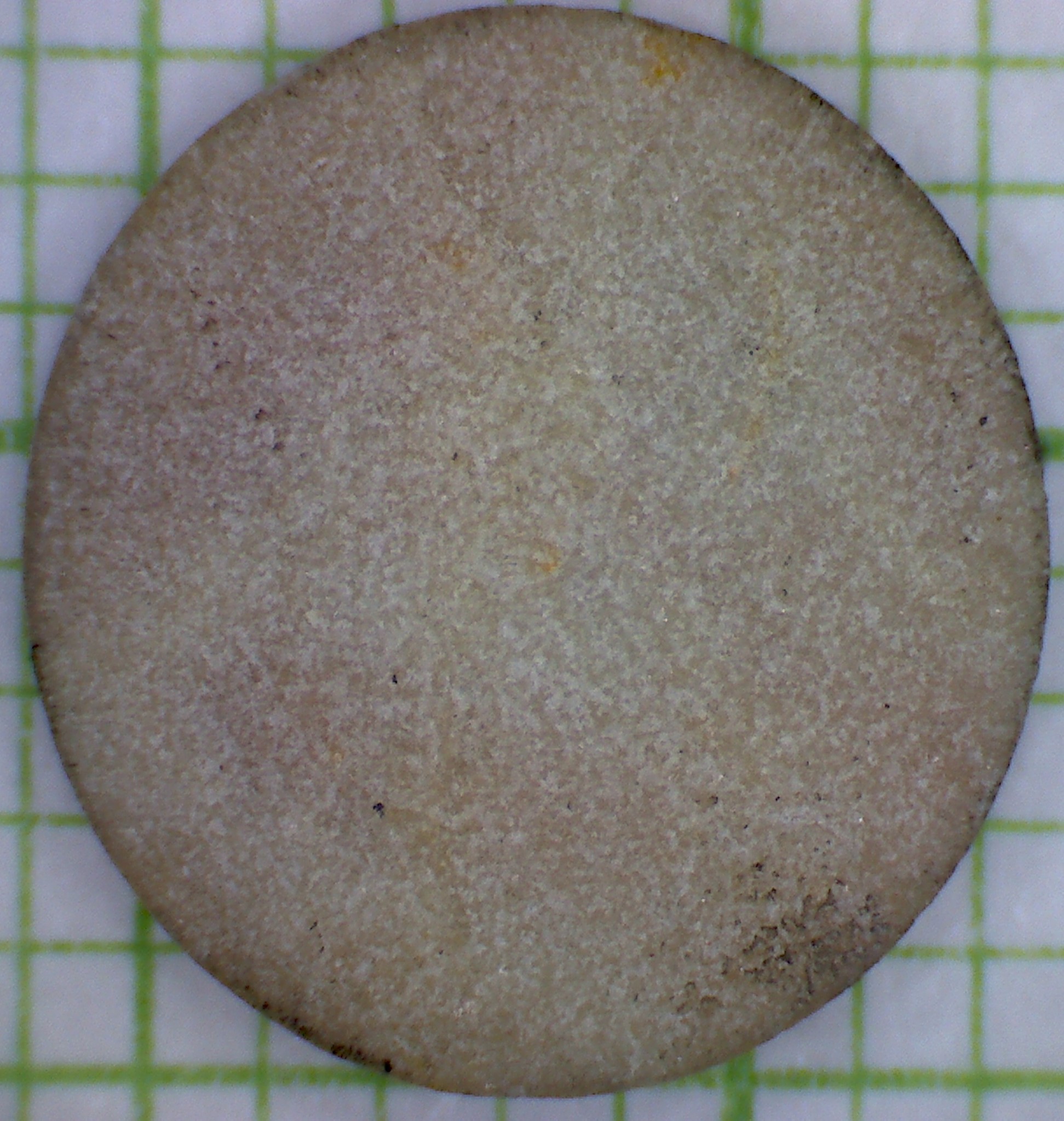}
	\includegraphics[width=2cm,height=2cm]{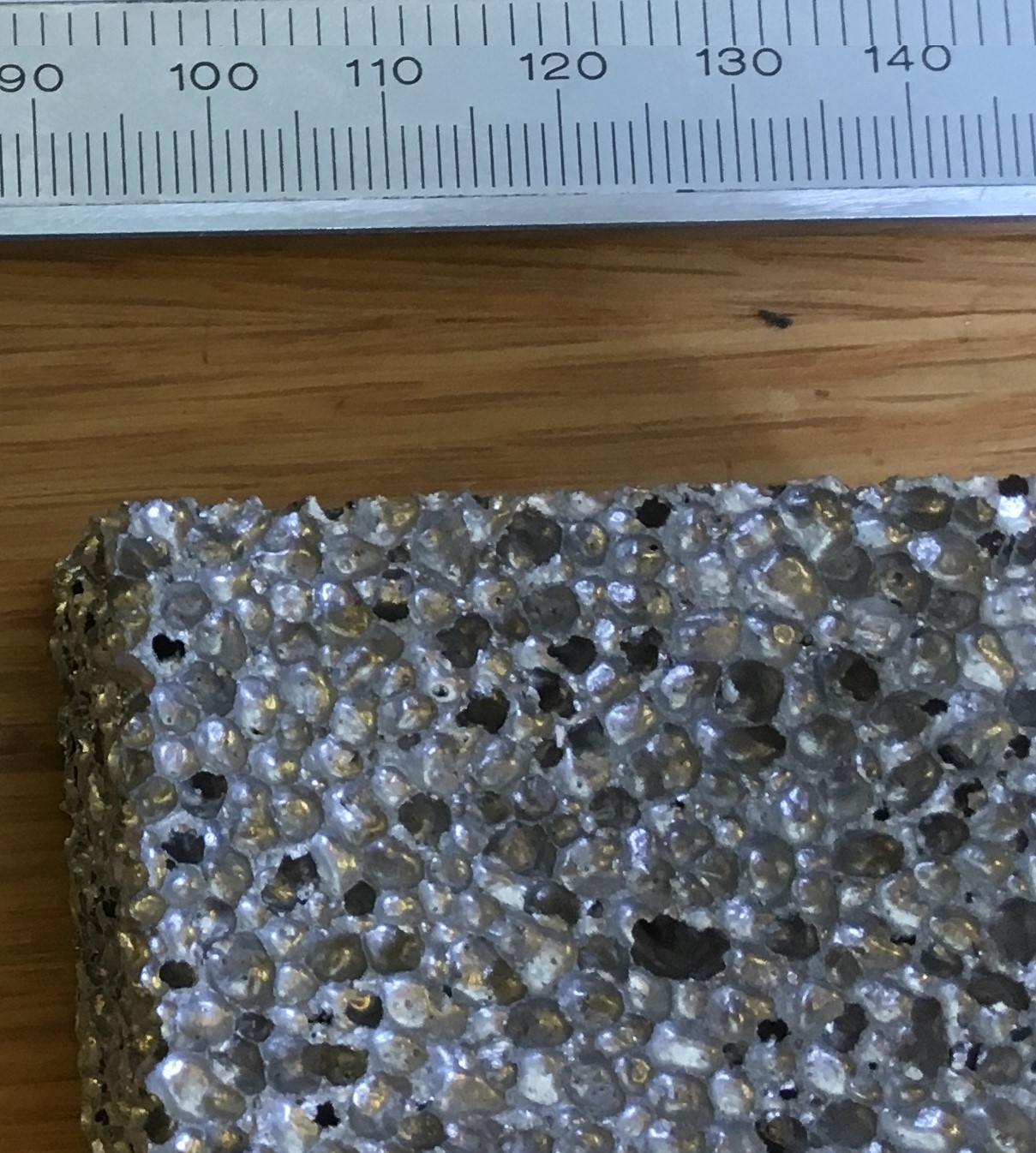}	
	\includegraphics[width=2cm,height=2cm]{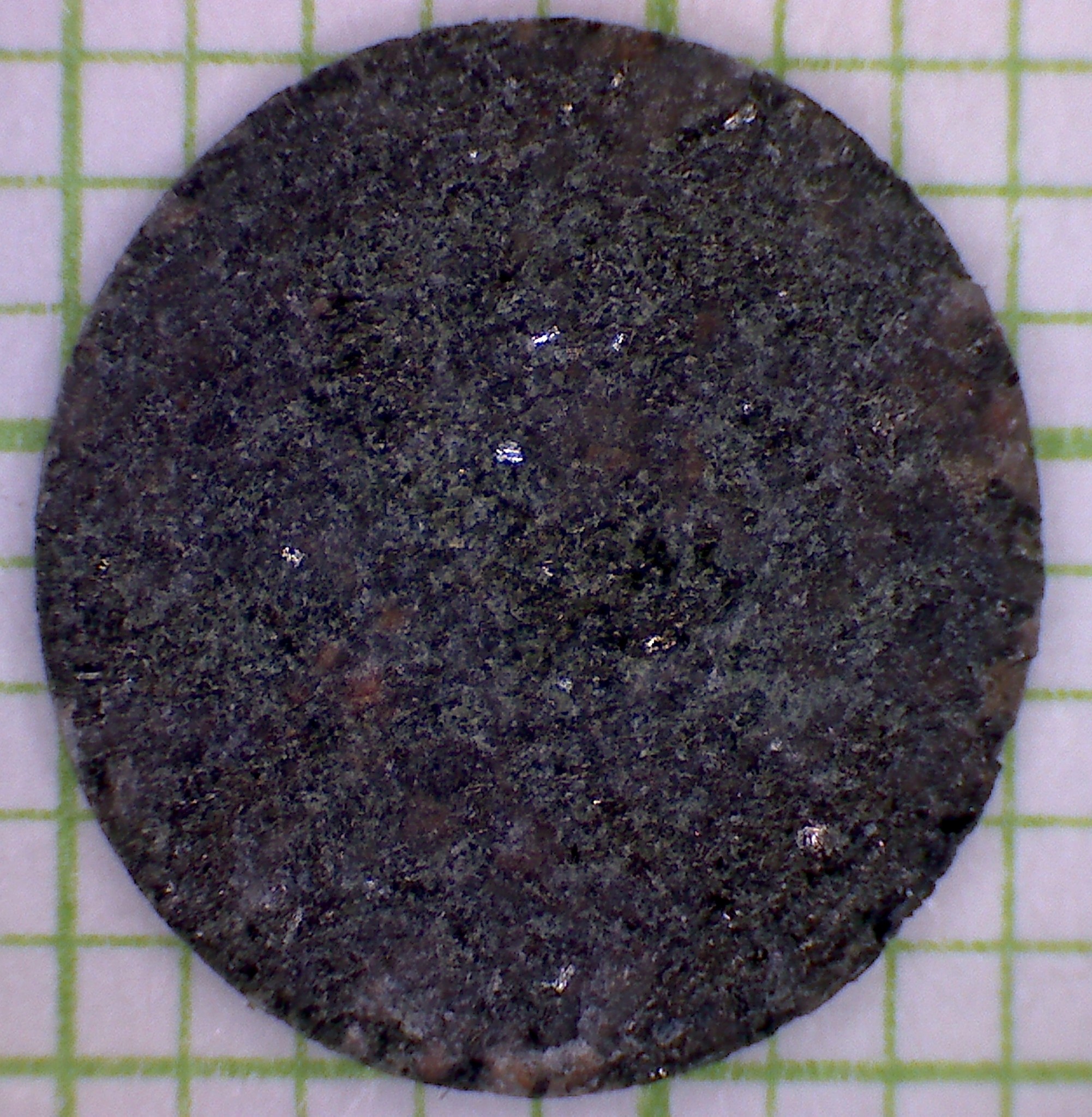}
	\caption{The heterogeneous structure of the measured materials. From left to right: capacitor, limestone from Vill\'any, metal foam, leucocrata rock with slires.}
\label{fig:spict}\end{figure}
\begin{figure}[ht]
	\centering
	\includegraphics[width=4cm]{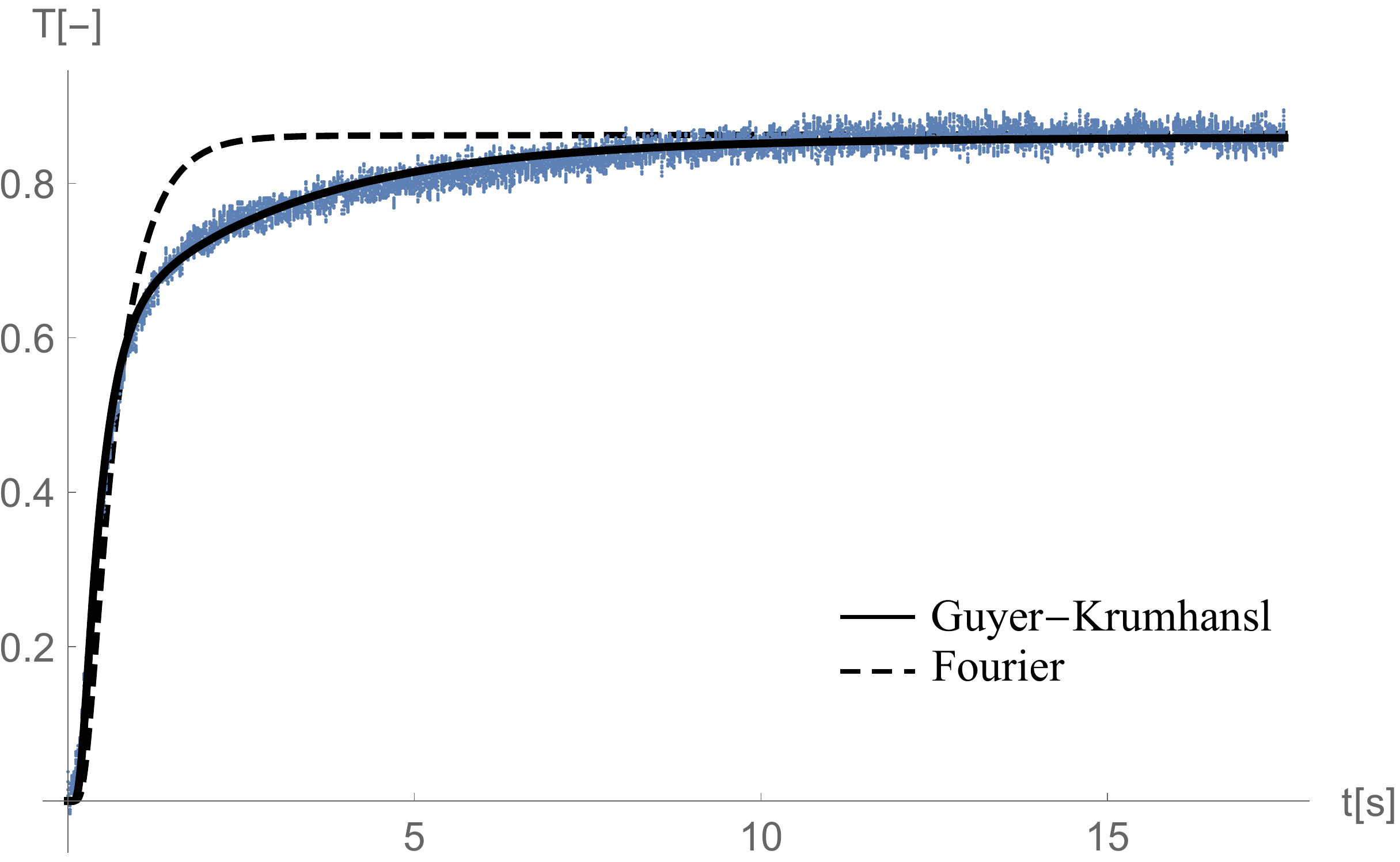}
	\includegraphics[width=4cm]{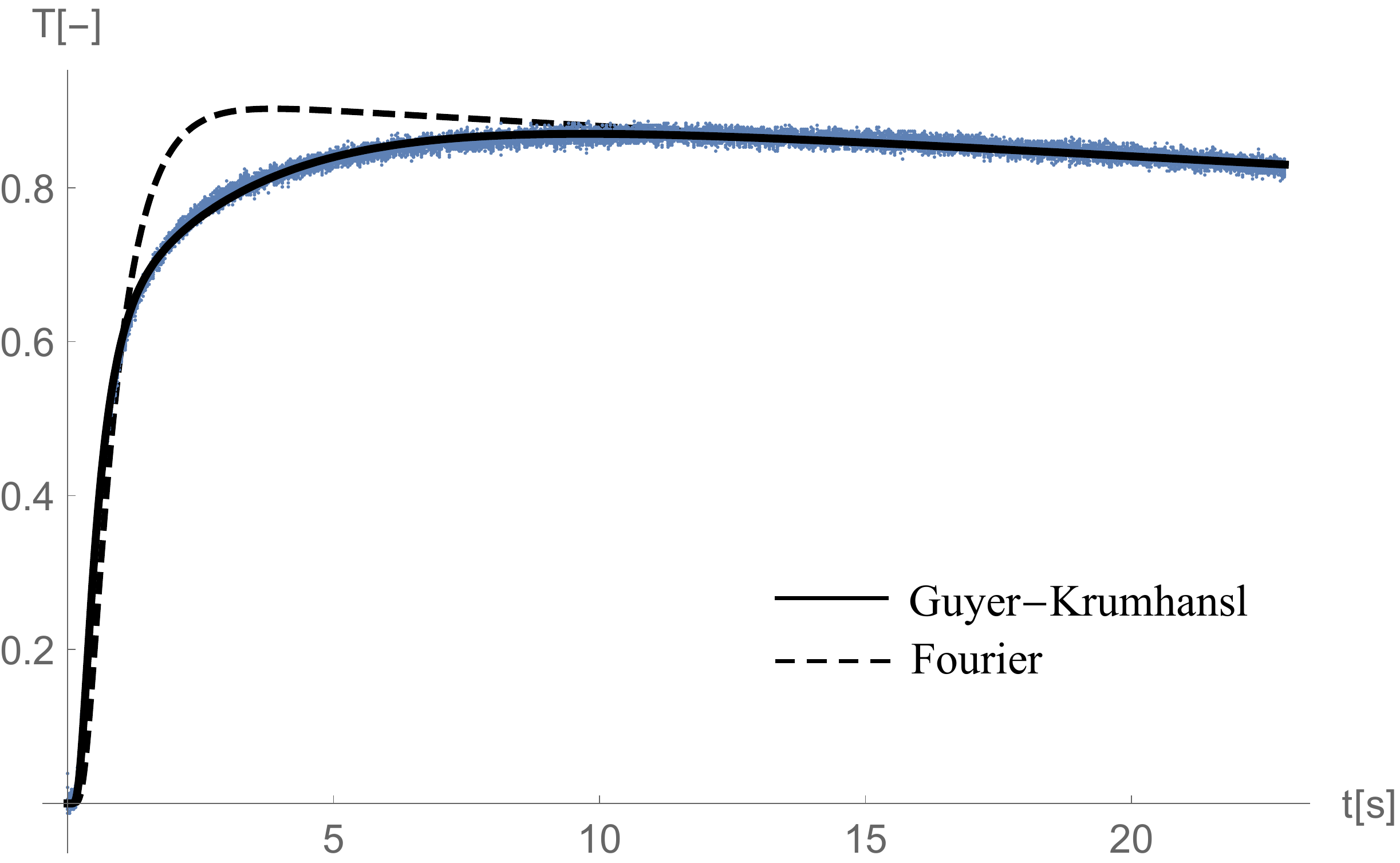}
	\caption{Left hand side: an artificial, layered capacitor sample, right hand side: limestone from Vill\'any (southern Hungary). The  experimentally measured data of the backside temperature as the function of time are in blue. The solid line is the best-fitted solution of the Guyer-Krumhansl equation, the dashed line is the best-fitted Fourier equation with correct asymptotics. }
\label{fig:fit-data12}\end{figure}
\begin{figure}[ht]
	\centering
	\includegraphics[width=4cm]{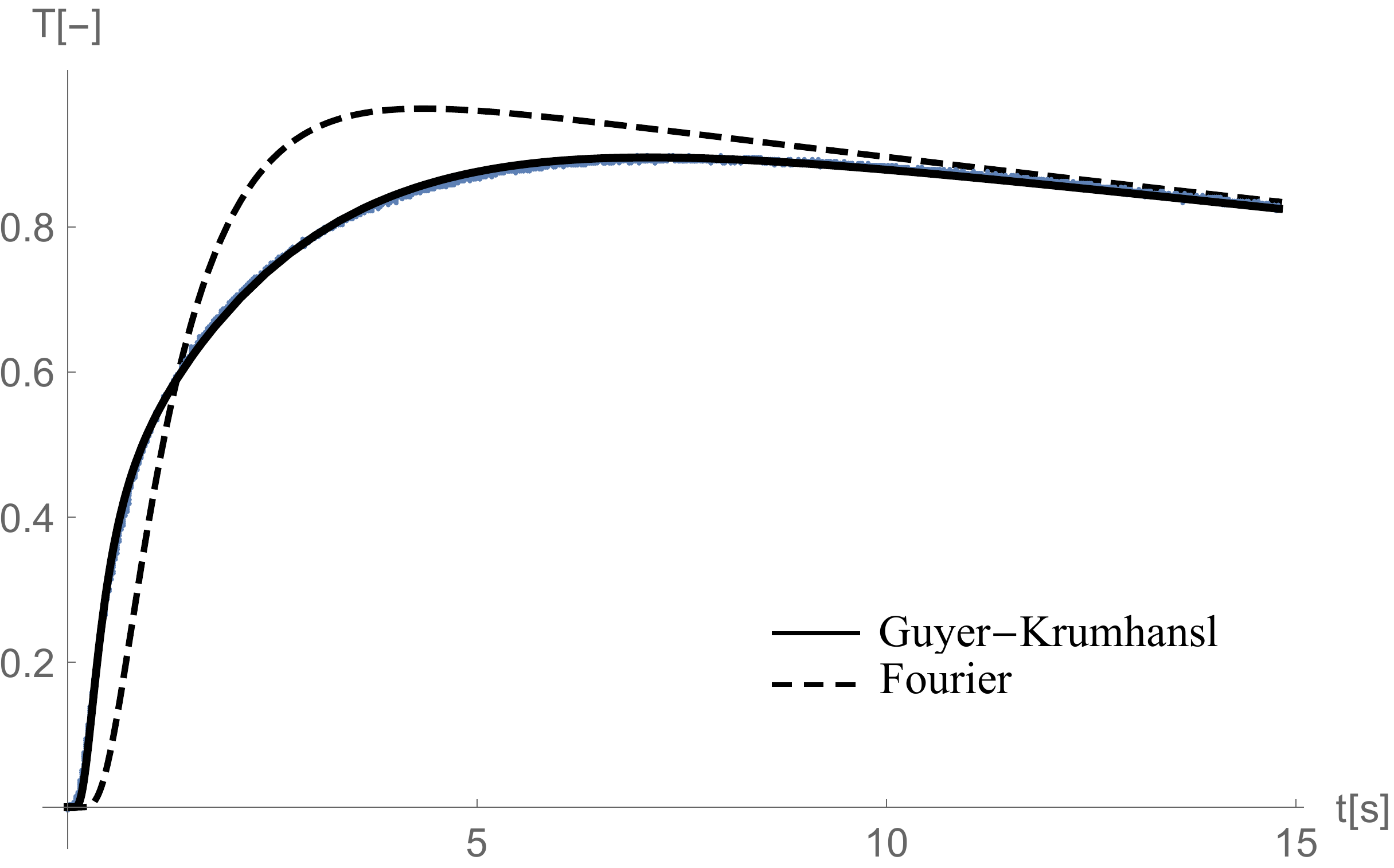}
	\includegraphics[width=4cm]{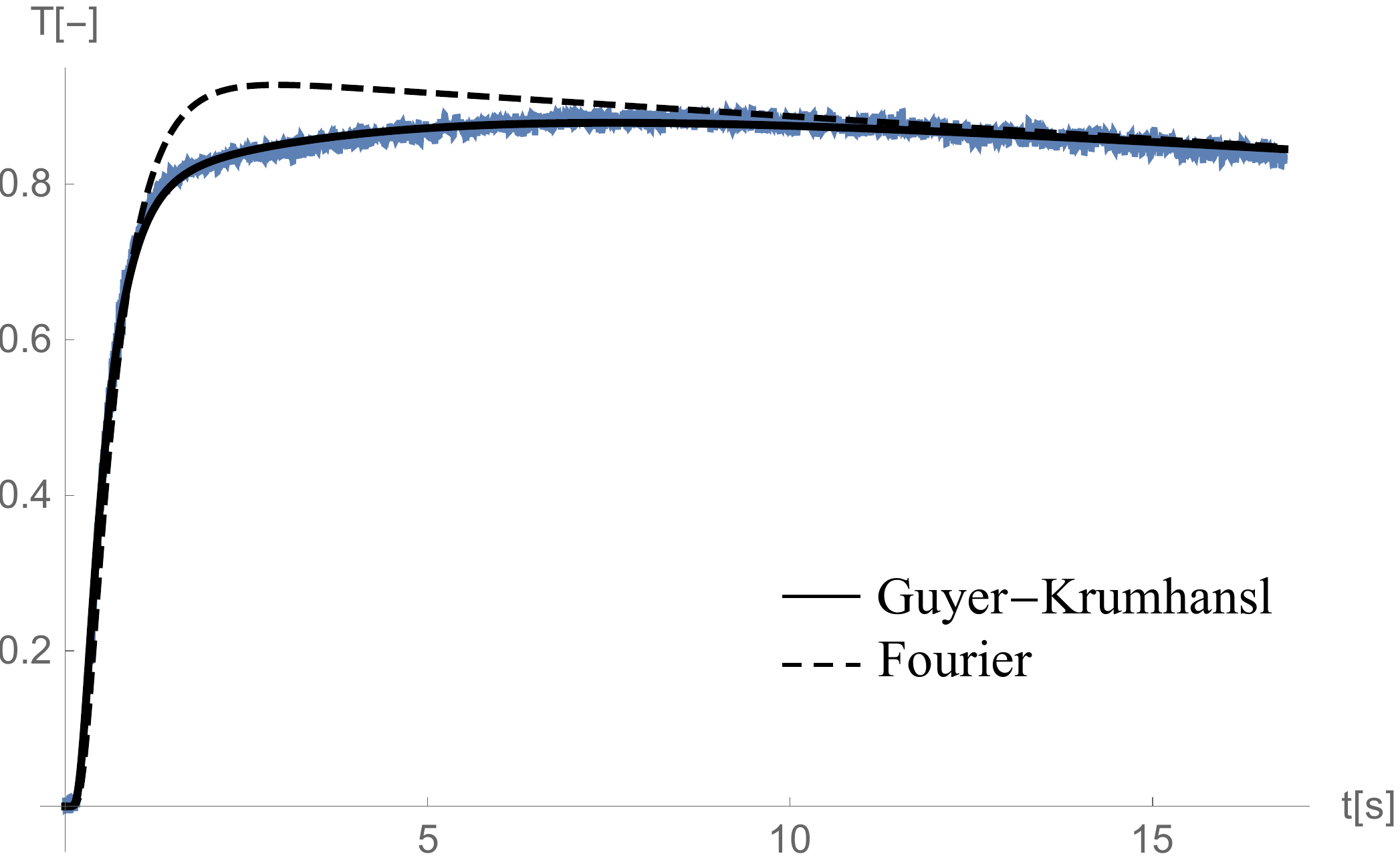}
	\caption{Left-hand side: metal foam sample 1, right-hand side: Leucocrata rock with slires. The  experimentally measured data of the backside temperature as the function of time are in blue. The solid line is the best-fitted solution of the Guyer-Krumhansl equation, the dashed line is the best-fitted  Fourier equation with correct asymptotics. }
\label{fig:fit-data34}\end{figure}

Among the investigated samples four had been chosen, where the deviation from Fourier heat conduction was the most apparent. These were a capacitor, the same as in \cite{BotEta16a}, a crystalline limestone sample from Vill\'any, southern Hungary, a leucocratic rock sample and a metal foam sample (see Figure \ref{fig:spict}.). The corresponding backside temperature is shown in Figures \ref{fig:fit-data12}. and \ref{fig:fit-data34}. The system of equations \re{ndebal}--\re{ndGK} are solved and the parameters are fitted to the data with the built-in  nonlinear regression algorithm of Mathematica 10.0, using the solution of the system of partial differential equations as an input function. On the figures the nondimensional temperature is scaled approximately to the adiabatic limiting temperature ($0.9 T_{max}$), the data is slightly smoothed with 3-point running average. The solid lines show the best GK fit and the dashed line is the best Fourier fit with the correct asymptotics.
The dimensionless parameters are given in Table \ref{nodpars}., and the related physical parameters in Table \ref{dpars}.
\begin{table}[h]
	\centering
	\begin{tabular}{l|cccc}
		Sample        &  $\hat\alpha\times 10^3$ & $\hat\tau$ & b \\\hline
		Capacitor     & 1.40  & 95.4 & 2.23 \\ 
		Limestone     & 1.124 & 99.1 & 2.17\\ 
		Metal foam 1  & 0.912 & 40.2 & 3.04 \\
		Leucocratic   & 1.56  & 132.0 & 1.77
	\end{tabular}
	\caption{Dimensionless parameters}\label{nodpars}
\end{table}
\begin{table}[h]
	\centering
	\begin{tabular}{l|ccccc}
		\ \ Sample & $L[mm]$ & $10^6\alpha_{F} [\frac{m^2}{s}]$ & $10^6\alpha_{GK}[\frac{m^2}{s}]$ & $\tau [s]$ & l[mm]\\\hline
		Capacitor     & 3.9  & 3.45 & 2.13    & 0.954 & 2.79 \\
		Limestone     & 1.7  & 0.45 & 2.950   & 0.991 & 0.80 \\
		Metal foam 1  & 5.1  & 3.04 & 2.373   & 0.402 & 1.70 \\
		Leucocratic    & 1.75 & 7.14 & 4.77    & 1.32  & 1.06 \\
	\end{tabular}
	\caption{Dimensional parameters. $\alpha_F$ is the thermal diffusivity from Fourier equation (dashed line), $\alpha_{GK}$ is the thermal diffusivity with the Guyer-Krumhansl equation (solid line)}
\label{dpars}
\end{table}
All measurements indicate an overdiffusive, $b>1$, Guyer--Krumhansl-type heat conduction.

\begin{figure}[ht]
	\centering
	\includegraphics[angle=270,width=4cm]{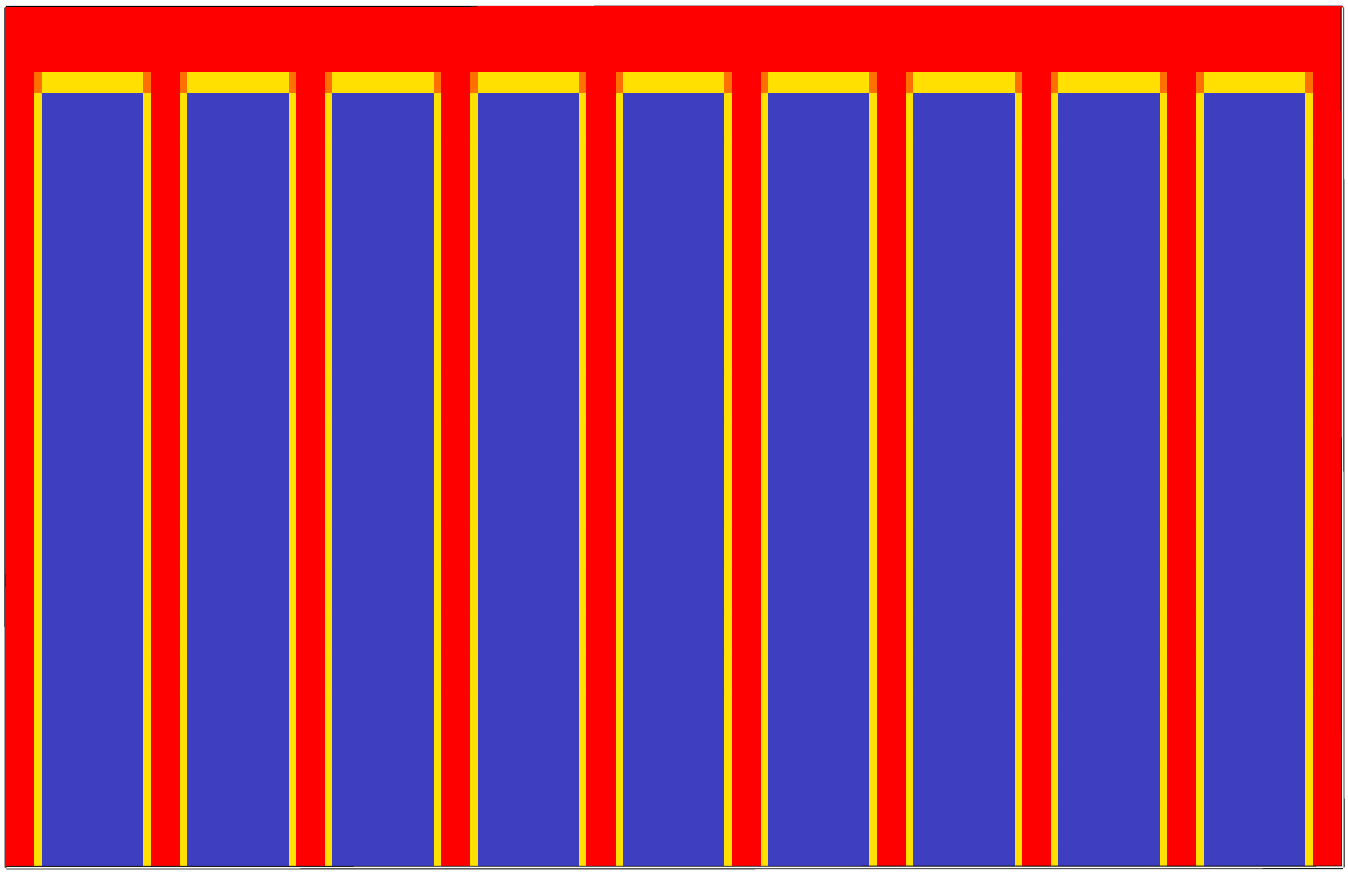}
	\caption{Geometry of the sample. The red color is Al, the blue is polystyrol. }
\label{fig:sketch2D}\end{figure}

In case of the capacitor sample the effect of regular heterogeneity can be calculated directly. The geometry of the calculation is shown in Figure \ref{fig:sketch2D}. It is assumed, that the cylindrical symmetry is not important and also an aluminium layer is applied at the upper (rear) side instead of the silver coating. The original capacitor is not symmetric, the aluminium and polystyrol layers are rolled and continuous, therefore, a two-dimensional representation seems to be proper approximation. 
The applied material parameters are given in Table \ref{mpars}.
\begin{table}[h]
	\centering
	\begin{tabular}{l|cccc}
		\ \  & Density & Heat capacity & Conductivity & Diffusivity  \\
		\ \  &  [$kg/m^2$] &  [$J/kg K$] & [$W/m K$] & [$m^2/s$]\\\hline
		Aluminium     & 2707  & 905  & 237    & 9.61$\times 10^{-5}$  \\
		Polystyrol    & 1040  & 1350 & 0.15   & $10^{-7}$  
	\end{tabular}
	\caption{Material parameters of the direct calculation}
	\label{mpars}
\end{table}

\begin{figure}[ht]
	\centering
	\includegraphics[angle=270,width=4cm]{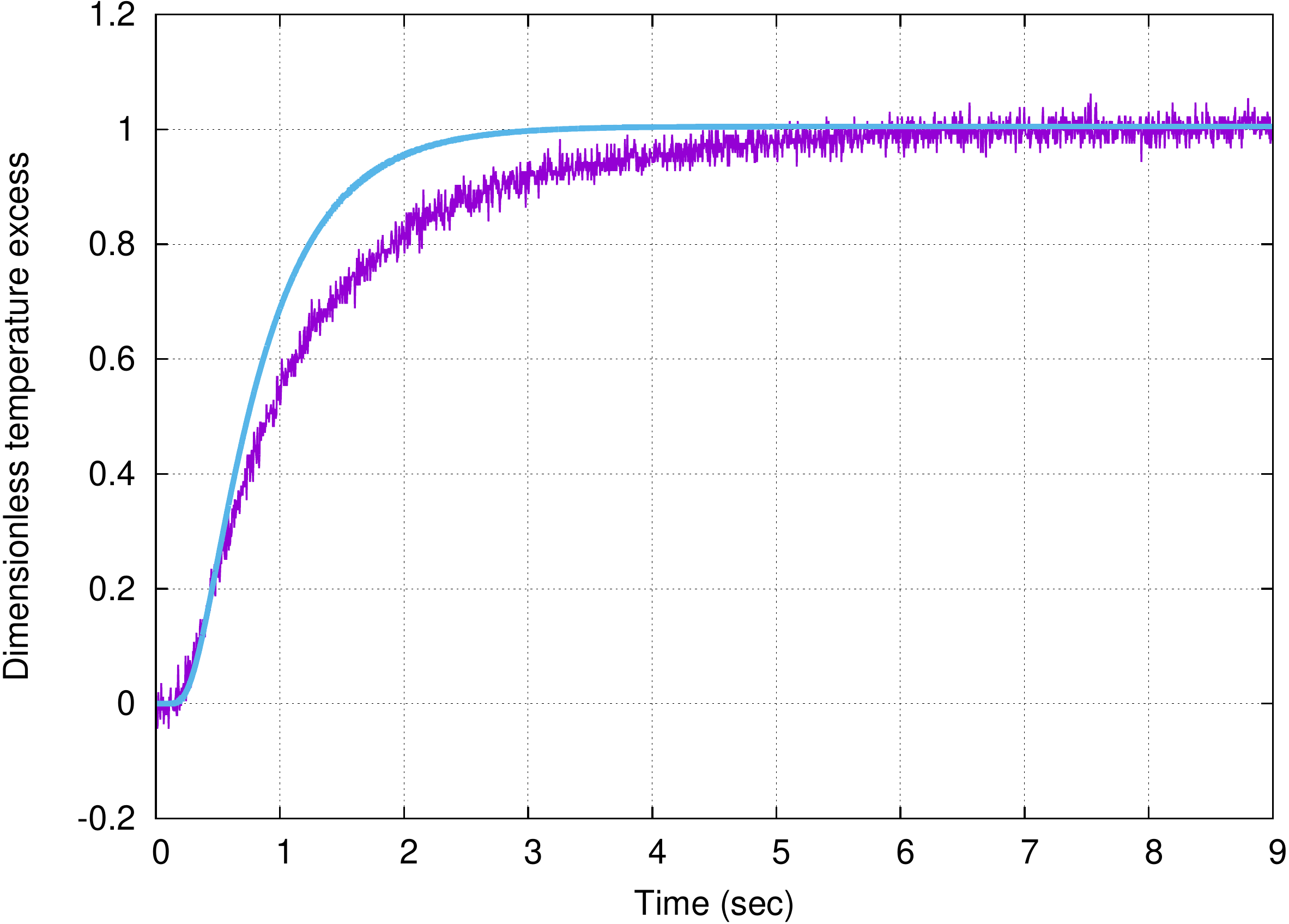}
	\includegraphics[angle=270,width=4cm]{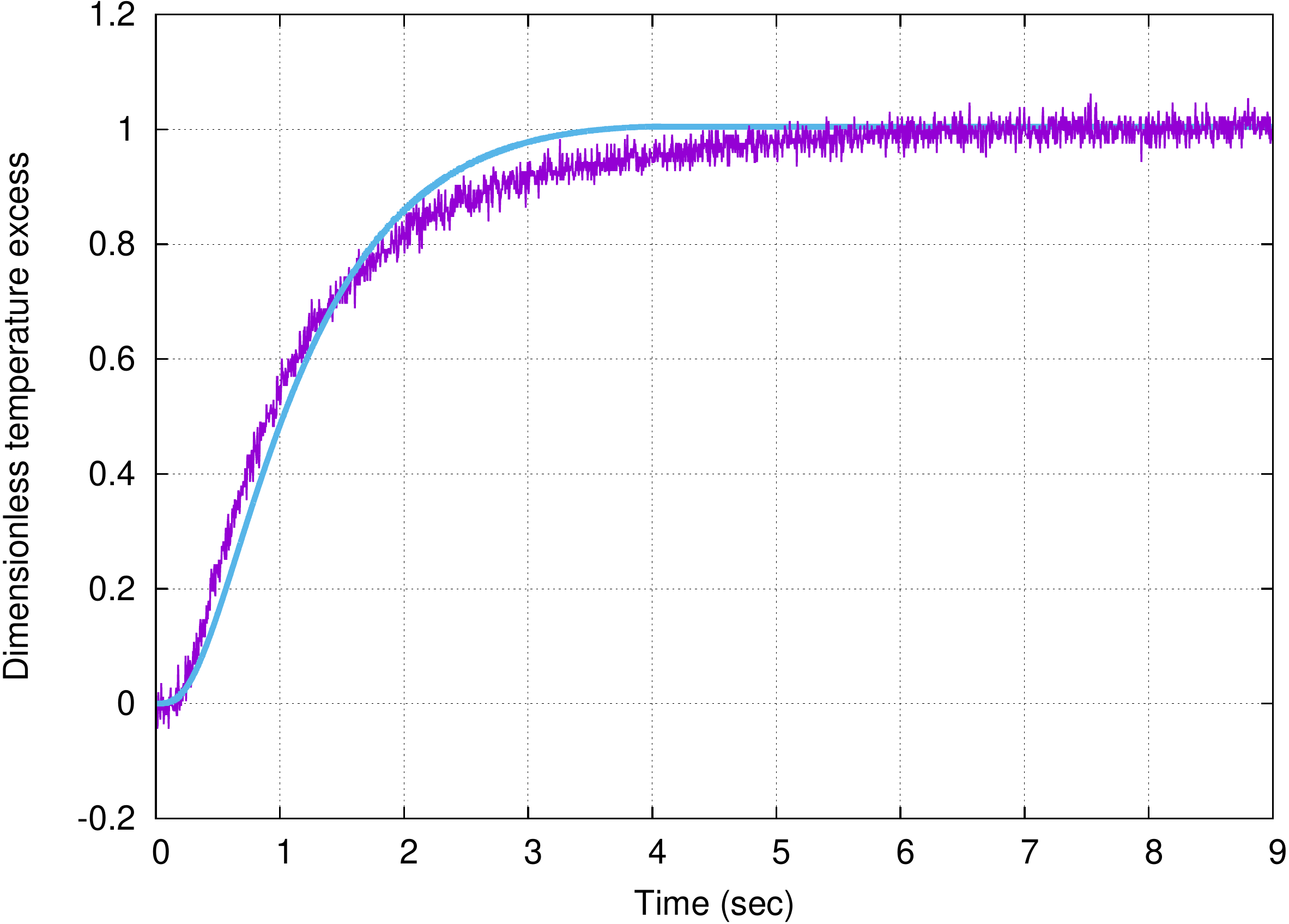}
	\caption{Micsrostructure simulation for the capacitor sample. Left: pure aluminium, right: heterogeneous aluminium-polystyrol two-dimensional simulation compared to the measured data.}
\label{fig:fit-datas1}\end{figure}

The Fourier equation (\re{hGK} with $\tau=0$) is solved numerically with a finite volume algorithm developed to deal with  material heterogeneity \cite{BerEta++}. The time step was chosen $\Delta t = 0.01 t_p = 10^{-4} s$ and the spatial dimensions were $185\times 40$ space steps, 5 and 15 steps for the $5\mu m$ aluminium and for the $15 \mu m$ polystyrol, respectively. The calculations are performed for $10s = 100000$ time steps in two cases. In the first case a pure aluminium sample was assumed and the comparison with the data of Both et al. \cite{BotEta16a} is shown on the left side of Figure \ref{fig:fit-datas1}. The measured and simulated dimensionless backside temperature is shown as a function of time by the thin and thick solid lines, respectively. With the geometry of Figure \ref{fig:sketch2D}. the material parameters of Table \ref{mpars}., the backside temperature is shown on the right side of Figure \ref{fig:fit-datas1}. 

One can see that for pure aluminium, after an initial coincidence, the slope of the computed curve becomes very different from the experimental values. For the heterogeneous case the simulated curve is closer to the experimental one, but the difference still exist.

We have shown deviation from Fourier's law in room temperature experiments. These were correctly modelled by the GK equation, known to be valid in case of low temperature solids. Our experimental samples exhibit heterogeneity. However, in case of the simplest, regular heterogeneity a direct simulation does not emulate the effect. These observations together indicate the robust universality of the GK equation in accordance with the theoretical expectations \cite{VanFul12a,KovVan15a}.
 
A practical use of our findings is the possibility to improve technological data of transient and stationary heat conduction in heterogeneous structures. This can simplify and guide microstructure based detailed analyses and computations.  

The work was supported by the grants NKFIH K104260, K116197 and K116375. Thanks to L\'aszl\'o Kov\'acs and the K\H omer\H o Kft. for the preparation of the rock sample. Thanks to Tam\'as B\'arczy and the Admatis Kft. for providing the metal foam sample.
\bibliographystyle{unsrt}

\end{document}